\documentclass[12pt,preprint]{aastex}

\newcommand{\msun}{$M_{\odot}\ $}

\newcommand{\lsim}{\raisebox{-0.3ex}{\mbox{$\stackrel{<}{_\sim} \,$}}}

\def\gta{\ifmmode {\mathbin{\lower 3pt\hbox   
    {$\,\rlap{\raise 5pt\hbox{$\char'076$}}\mathchar"7218\,$}}}
    \else {${\mathbin{\lower 3pt\hbox
    {$\rlap{\raise 5pt\hbox{$\char'076$}}\mathchar"7218\,$}}}
    $}\fi}
\def\lta{\ifmmode {\,\mathbin{\lower 3pt\hbox   
    {$\,\rlap{\raise 5pt\hbox{$\char'074$}}\mathchar"7218\,$}}}
    \else {${\mathbin{\lower 3pt\hbox
    {$\rlap{\raise 5pt\hbox{$\char'074$}}\mathchar"7218\,$}}}
    $}\fi}

\shorttitle {IMPLICATIONS OF ATOMIC LINES FROM THE SURFACES OF
NEUTRON STARS} \shortauthors {Bhattacharyya, Miller, \& Lamb }

\begin{document}

\title {THE SHAPES OF ATOMIC LINES FROM THE SURFACES OF WEAKLY
MAGNETIC ROTATING NEUTRON STARS AND THEIR IMPLICATIONS}

\author {Sudip Bhattacharyya\altaffilmark{1}, M. Coleman
Miller\altaffilmark{1,2}, and Frederick K. Lamb\altaffilmark{3,4}}

\altaffiltext{1}{Department of Astronomy, University of Maryland at
College Park, College Park, MD 20742-2421}

\altaffiltext{2}{Also Maryland Astronomy Center for Theory and
Computation}

\altaffiltext{3}{Center for Theoretical Astrophysics, University of
Illinois at Urbana-Champaign, Loomis Laboratory of Physics, 1110
West Green Street, Urbana, IL 61801-3080}

\altaffiltext{4}{Also Departments of Physics and Astronomy,
University of Illinois at Urbana-Champaign}

\begin{abstract}

Motivated by the report by Cottam et al.\ (2002) of iron resonance
scattering lines in the spectra of thermonuclear bursts from
\hbox{EXO~0748$-$676}, we have investigated the information about
neutron star structure and the geometry of the emission region that
can be obtained by analyzing the profiles of atomic lines formed at
the surface of the star. We have calculated the detailed profiles of
such lines, taking into account the star's spin and the full effects
of special and general relativity, including light-bending and
frame-dragging. We discuss the line shapes produced by rotational
Doppler broadening and magnetic splitting of atomic lines, for the
spin rates and magnetic fields expected in neutron stars in low-mass
X-ray binary systems. We show that narrow lines are possible even
for rapidly spinning stars, if the emission region or the line of
sight are close to the spin axis. For most neutron stars in low-mass
systems, magnetic splitting is too small to obscure the effects of
special and general relativity. We show that the ratio of the star's
mass to its equatorial radius can be determined to within 5\% using
atomic line profiles, even if the lines are broad and skewed. This
is the precision required to constrain strongly the equation of
state of neutron star matter. We show further that if the radius and
latitude of emission are known to $\sim 5-10$\% accuracy, then
frame-dragging has a potentially detectable effect on the profiles
of atomic lines formed at the stellar surface.

\end{abstract}


\section {Introduction} \label{sec: 1}

The discovery (Cottam, Paerels, \& M\'endez 2002) of line features
in the average spectrum of 28 type-I X-ray bursts from the neutron
star in the low-mass-X-ray-binary (LMXB) \hbox{EXO~0748$-$676} is a
significant advance that provides new information on the structure
of neutron stars and the properties of dense matter. The lines were
detected using XMM-Newton and have been identified as resonance
scattering lines of Fe~XXV H$\alpha$, Fe~XXV 2--3, and possibly
O~VIII Ly$\alpha$. The gravitational redshifts inferred from the
observed energies of all three lines are $\sim$\,0.35 (Cottam et
al.\ 2002), consistent with their formation at the surface of the
star.

The profiles of atomic lines formed at the surface of a neutron star
are affected not only by the gravitational redshift there, but also
by the star's radius, spin, angular momentum, and magnetic field.
Such lines carry more information about the properties of the star,
the equation of state (EOS) of cold, high-density matter, and the
binary system than any other features of the star's spectrum.

The X-ray burst oscillations of \hbox{EXO~0748$-$676} show that the
spin frequency of the neutron star in this system is 45~Hz
(Villarreal \& Strohmayer 2004), implying a line-of-sight rotational
speed at the stellar surface $\lsim 0.015$~c and a rotational
Doppler width at most ten times the Fe line width produced by Stark
broadening (see Bildsten, Chang, \& Paerels 2003), even for the high
system inclination $i\sim 75^{\rm o}$--$82^{\rm o}$ indicated by the
dips and eclipses observed in this system (Parmar et al.\ 1986).

Spin rates as low as 45~Hz
are expected only if the accretion phase has
lasted less than the time required to spin up the star, or the
accretion rate is low enough or the star's magnetic field strong
enough that the star has reached spin equilibrium (Lamb \& Yu 2004).
Magnetic splitting by fields $\gta 2 \times 10^{9}$~G would change
the profiles of the Fe lines significantly, and fields an order of
magnitude stronger would shift the energy $E_{0}$ of the line center
(Loeb 2003). Consequently, the strength of the magnetic field in
\hbox{EXO~0748$-$676} and in other burst sources is an important
issue.

Some 16 neutron stars in LMXBs have spin rates ranging from 200~Hz
to 600~Hz (see Lamb \& Yu 2004), much higher than the spin rate of
the neutron star in \hbox{EXO~0748$-$676}. Spin rates this high
imply surface speeds near the rotational equator of 0.05--0.20~c.
Atomic lines formed at the surfaces of such rapidly spinning stars
may be broadened substantially by rotational Doppler broadening
(\"Ozel \& Psaltis 2003). The magnitude of this broadening is an
important issue in determining whether general relativistic effects
can be detected.

Here we summarize the most important results of a detailed
theoretical study of the effects of general relativity on the shapes
of atomic spectral lines formed at the surfaces of spinning neutron
stars. Some of these effects were illustrated by \"Ozel \& Psaltis
(2003). We also discuss the evidence concerning the strengths of the
magnetic fields of neutron stars in LMXBs and the implications for
magnetic splitting of lines formed near their surfaces and for
measurement of general relativistic effects.

We find that even lines formed at the surfaces of stars with spin
rates as large as hundreds of Hertz observed at inclinations to the
spin axis as high as $\sim 75^{\rm o}$ can have relative widths
${\delta E}_0/E_0$ as small as $10^{-2}$, if they are formed
predominantly within $\sim$\,10$^\circ$--20$^\circ$ of the spin
axis. We show that using the geometric mean of the energies of the
line wings greatly reduces the systematic error in determinations of
the star's mass-to-radius ratio. We also show that the ratios of the
depths of the red and blue flux minima of a resonance scattering
line provide a unique signature of frame-dragging. In most neutron
stars in low-mass systems, magnetic splitting of lines is too small
to obscure the effects of special and general relativity on their
profiles. Finally, we discuss the prospects for detecting these
relativistic effects. In the rest of this paper we use geometric
units in which $c=G=1$, unless otherwise stated.

\section {Physical Effects and Calculational Method} \label{sec: 2}

At a given instant, the X-ray emitting portion of the neutron star
surface could be a hot spot (produced, e.g., by magnetically
channeled accretion, confined thermonuclear burning, or localized
cooling gas), a belt (produced, e.g., by disk accretion near the
rotational equator or rotationally confined thermonuclear burning),
or a combination. Emission from a localized region that is not at
the rotation pole is effectively smeared into an axisymmetric belt
when averaged over many spin periods. This is to be expected for the
spin frequencies $\gta10$~Hz of neutron stars in LMXBs and the
integration times of seconds typical of current high-resolution
spectroscopic measurements. We therefore treat the emission as
coming from an axisymmetric belt and assume for simplicity that the
belt is uniformly bright.

The profile of a spectral line formed at the stellar surface carries
the imprint of several physical effects. Here we consider
(1)~magnetic (Zeeman or Paschen-Back) splitting by the star's
magnetic field, (2)~longitudinal and transverse Doppler shifts,
(3)~special relativistic beaming, (4)~gravitational redshifts,
(5)~light-bending, and (6)~frame-dragging. We treat the star as
spherical, neglecting the effects of the spin-induced stellar
oblateness and the deviation of the exterior spacetime from Kerr,
which are small compared to the uncertainties.\footnote {For a
stellar model with a gravitational mass $\ge 1.4M_\odot$ and a spin
rate of 400~Hz as measured at infinity, constructed using the
A18$+\delta v+$UIX EOS (Akmal, Pandharipande, \& Ravenhall 1998),
the ratio of the polar radius to the equatorial radius is greater
than 0.965; all other effects of the star's oblateness are at most a
few percent. The Kerr spacetime is an accurate approximation to the
exterior spacetime of a spinning neutron star only to first order in
the spin; higher accuracy requires numerical computation of the
spacetime (see Miller, Lamb, \& Cook 1998).}

Effects~(1)--(6) influence the line shape in different ways.
Magnetic splitting, Doppler shifts and the variation in the emitting
area for a given observed energy cause an otherwise narrow
scattering line to become broader and can cause it to have two
minima. The gravitational redshift and transverse Doppler effect
both shift the line toward lower energies. Special relativistic
beaming deepens the blue minimum but suppresses (or eliminates) the
red minimum, for emission from a belt around the star. Light-bending
deepens both minima, but deepens the red minimum more and can
sometimes make visible a red minimum that was suppressed by special
relativistic beaming. Finally, frame-dragging broadens the line and
shifts it slightly toward higher energies.

To compute detailed line shapes, we first assumed (1)~an EOS, (2)~a
gravitational mass $M$, and (3)~a stellar spin rate $\nu_{s}$. We
chose as a representative modern\footnote {Models that, like 
A18+$\delta$v+UIX,
fit the Nijmegen database (Stoks et al.\ 1993) with $\chi^2/N_{\rm
data} \sim 1$ are called `modern'.} EOS the A18$+\delta v+$UIX model
of Akmal et al.\ (1998), which combines the Argonne $v_{18}$ (A18)
model of the two-nucleon-interaction (Wiringa, Stoks, \& Schiavilla
1995), the Urbana IX (UIX) model of the three-nucleon interaction
(Pudliner et al.\ 1995), and relativistic boost corrections to the
A18 interaction. Using this EOS, we determined the star's radius $R$
and dimensionless angular momentum $j \equiv J/M^2$ by computing
numerically and self-consistently the structure and interior
spacetime of the star, following Cook, Shapiro, \& Teukolsky (1994;
see Bhattacharyya et al.\ 2000 for a detailed description of the
method we used).

We usually consider an emitting belt in the northern hemisphere
specified by (4)~the colatitude $i_{\rm belt}^{\rm p}$ of its
northernmost (polar) edge and (5)~the colatitude $i_{\rm belt}^{\rm
eq}$ of its southernmost (equatorial) edge. Alternatively, we
specify the location and width of the belt by the colatitude $i_{\rm
belt}$ of its center and its angular half-width $\Delta i_{\rm
belt}$. The emission from each point in the belt is specified in the
corotating frame by (6)~an intensity distribution of the form
$I_\nu\propto \cos^n(\psi)$, where $\psi$ is the angle measured from
the local normal to the stellar surface and the parameter $n$
specifies the degree of beaming. Except when we consider magnetic
splitting, we assumed a boxcar intrinsic line shape with (7)~a
halfwidth ${\delta E}_0 = 10^{-3} E_0$, where $E_0$ is the rest
energy of the line center; this is comparable to the width expected
from Stark broadening (Bildsten et al.\ 2003). The final parameter
is (8)~the inclination $i_{\rm spin}$ of the observer relative to
stellar spin axis.

Given these eight parameters, we determine the image of the source
seen by the observer by tracing the path of photons in the Kerr
space-time from the observer back to the source. We then compute the
observed spectrum by integrating the energy-dependent specific
intensity over the image (for a more detailed description of this
procedure, see Bhattacharyya, Bhattacharya, \& Thampan 2001). Our
method reproduces (for $j = 0.0)$ the results of \"Ozel \& Psaltis for their
parameter choices. The calculations reported here ignore
instrumental broadening, to show what may be possible in the future.

\section {Results} \label{sec: 3}

The combined effects of magnetic splitting, Doppler shifts, special
relativistic beaming, gravitational redshift, light-bending, and
frame-dragging typically make lines broader, shallower, and skewed,
and can create two flux minima (see Fig.~1, where the profiles are
scaled to have the same equivalent width). These effects do not
weaken lines, contrary to what has sometimes been implied
in the literature (see, e.g., \"Ozel \& Psaltis 2003): their
equivalent widths remain unchanged. Note, however, that a 
broader and shallower line might be more difficult to detect.

Whether magnetic splitting is an important broadening mechanism
depends on the strength of the star's surface magnetic field.
Magnetic fields in the line-forming region with strengths $\gta
10^{10}$~G would broaden and shift X-ray lines enough to make
identification of the transitions and determination of the redshift
and other stellar properties difficult (Loeb 2003). However, the
spin rates and other observed properties of neutron stars in LMXBs
and the spin-down rates of recycled radio pulsars indicate that most
neutron stars in LMXBs have magnetic fields $\sim 10^7$---$10^9$~G
(Miller, Lamb, \& Psaltis 1998; Lamb \& Yu 2004), too weak to
broaden Fe lines by more than $\sim 20$~eV.

The spin rate of the neutron star in \hbox{EXO~0748$-$676} could be
45~Hz either because it has not accreted enough matter to reach a
higher frequency or because its field is strong enough to maintain
spin equilibrium at 45~Hz. Accretion will double the spin rate
$\nu_{s}$ of a star far from spin equilibrium in $\sim 6 \times 10^6
({\dot M}/0.03{\dot M}_E)^{-0.87} (\nu_{s}/45~{\rm Hz})$~yr, where
${\dot M}$ is the accretion rate and ${\dot M}_E \approx 1.5 \times
10^{18}$~g~s$^{-1}$ is the Eddington accretion rate onto a star with
a 10~km radius (see Lamb \& Yu 2004). The magnetic field required to
maintain a 45~Hz spin rate is $2 \times 10^9$~G for accretion at the
relatively high long-term average rate $\langle {\dot M} \rangle
\sim 0.03 {\dot M}_E$, or $1.5 \times 10^8$~G for $\langle {\dot M}
\rangle \sim 10^{-4} {\dot M}_E$ (Lamb \& Yu 2004). The splitting
produced by $2 \times 10^9$~G would be $\sim 20$~eV, so it is
unlikely that magnetic splitting dominates the line widths observed
in \hbox{EXO~0748$-$676}, but it could contribute to them.

A line produced at the surface of a star that is spinning rapidly
may nevertheless appear narrow if it is intrinsically narrow and
(a)~the colatitude $i_{\rm belt}^{\rm eq}$ of the region where the
line is formed is sufficiently small, (b)~the observer's inclination
$i_{\rm spin}$ relative to the spin axis is sufficiently small, or
both angles are relatively small. For example, the observed width of
Fe lines formed at the surface of a $1.5$\msun star observed at
$i_{\rm spin} \approx 75^\circ$ is only $\sim 20$~eV if $\nu_{s} =
200$~Hz and $i_{\rm belt}^{\rm eq} \lta 24^{\rm o}$, or if $\nu_{s}
= 600$~Hz and $i_{\rm belt}^{\rm eq} \lta 7^{\rm o}$. Thus, narrow
lines are possible even if $\nu_{s} = 600$~Hz, if they are
preferentially formed near the rotation axis, as in a hotter region
close to a nearly aligned magnetic pole (in some evolutionary
scenarios, alignment of the magnetic axis with the spin axis is
expected).

Scattering lines with two flux minima are possible for emission from
any part of a spherical surface only if the emission is beamed in
the rest frame of the surface or there is general relativistic light
deflection; the depths of the resulting minima depend on the
location and width of the emitting belt and $i_{\rm spin}$.

\"Ozel \& Psaltis (2003) showed that if the redshift $z_{\rm surf}$
at the surface of the star is estimated using the energy of the
deepest minimum in the full line profile, the distortion of the line
profile by special and general relativistic effects can introduce
errors of tens of percent for stars with spin rates of hundreds of
Hz. Our study confirms this result (see Fig.~2).

We have found a way to reduce greatly the effects of the distortion
of the line profile on measurements of $z_{\rm surf}$. \textit
{Using the quantity $(E_0/E_{\rm gm})-1$, where $E_{\rm gm} \equiv
(E_1E_2)^{1/2}$ is the geometric mean of the low-energy and
high-energy edges of the observed line, yields very accurate
estimates of $z_{\rm surf}$, even if the line is broad and skewed.}
We explored this approach because, for surfaces rotating at speed
$\beta$ toward and away from the observer, the lower- and
higher-energy edges of the line are Doppler shifted to
$E_1=E_0\gamma(1-\beta)$ and $E_2=E_0\gamma(1+\beta)$, where
$\gamma\equiv(1-\beta^2)^{-1/2}$ is the Lorentz factor. $E_{\rm gm}$
is therefore exactly equal to $E_0$ in this geometry, when
light-bending is neglected. Our calculations show that using the
geometric mean of the energies of the two flux minima also yields a
very accurate estimate of $z_{\rm surf}$.

Fig.~2 compares the value of $R/M$ estimated using $E_{\rm gm}$
with the true equatorial value of $R/M$ and the values estimated
using (a)~the geometric means of the energies of the lower- and
higher-energy flux minima and edges of the line, (b)~the arithmetic
means of the energies of lower- and higher-energy flux minima and
the energies of the lower- and higher-energy edges of the line, and
(c)~the energy of the deepest flux minimum in the line profile,
assuming that the line is formed in the region between colatitude
$60^{\rm o}$ and the rotational equator $90^{\rm o}$. The error in
$R/M$ is greater when $R/M$ is estimated from the energy of the
deepest flux minimum in a line profile formed over a range of
latitudes than in a line formed uniformly over the whole stellar
surface (see \"Ozel \& Psaltis 2003), but the former geometry is
probably more realistic.

The value of $R/M$ estimated using $E_{\rm gm}$ differs from the
ratio of the star's equatorial radius $R_{\rm eq}$ to $M$ by less
than 1.7\% for $\nu_{s} < 600$~Hz (see Fig.~2). Even if the line is
formed over the whole surface of the star, the geometric mean gives
a much better estimate of $R/M$ than does the energy of the deepest
flux minimum. Using the arithmetic mean of the energies of the two
flux minima or the energies or the energy of the deepest flux
minimum gives estimates of $R/M$ that differ from $R_{\rm eq}/M$ by
up to 7.3\% and 57.0\%, respectively.

The estimated value of $R/M$ deviates most from the true value when
the star spins rapidly, $R/M$ is small (for a fixed EOS model and 
stellar spin), or the band is
centered at the spin equator.  We find that the accuracy is
independent of the beaming parameter $n$.  In all of the
combinations of parameters we have tried, the accuracy in $R/M$
using our method is better than 2\% when $\nu_s=600$~Hz.  In
Figures~3 and 4 we show the dependence of the estimates on the true
values of $R/M$ and belt locations and widths.

Importantly, we have found a possible signature of frame-dragging in
the line profile, which could be used to detect frame-dragging if
$R\sin i_{\rm spin}$ is known to better than $\sim 5-10$\%. As pointed out by
the anonymous referee, the effect of frame-dragging for a given
angular velocity is to decrease the linear speed of the surface
relative to the locally nonrotating frame (also see Poutanen \&
Gierli\'nski 2003).  The angular velocity of frame dragging at the surface is
typically $\sim 10-20$\% of the stellar spin rate. 
As seen in Fig.~5, this affects  the ratio
of the depth of the low-energy (red) flux minimum to the depth of
the high-energy (blue) flux minimum, which increases steeply with
increasing $j$, but either decreases or increases only slowly with
variations of most other parameters.  This effect is also
produced by changing the linear surface speed at a fixed spin
frequency, i.e., by changing $R\sin i_{\rm spin}$. Because $R\sin i_{\rm spin}$ also
affects the width of the full line profile, a precise $(\sim 5-10$\%) measurement of
line-width (and hence
$R\sin i_{\rm spin})$ could allow detection of frame-dragging effects.

If such precision is possible (e.g., with a future high-throughput
and high-resolution mission such as Constellation-X), then the most
robust way to detect this effect is to
measure the overall asymmetry of the line,
using as a reference the local flux maximum $\phi_{m}$ at $E_{m}$,
between the two flux minima at $E_1$ and $E_2$. The photon flux is
equal to $\phi_{m}$ at two other energies, $E_{m1}<E_{m}$ and
$E_{m2}>E_{m}$, but less than $\phi_{m}$ at all energies between
$E_{m1}$ and $E_{m2}$. There is therefore a ``red flux deficit''
from $E_{m1}$ to $E_{m}$ relative to $\phi_{m}(E_{m} - E_{m1})$ and
a ``blue flux deficit'' from $E_{m}$ to $E_{m2}$ relative to
$\phi_{m}(E_{m2} - E_{m})$. Like equivalent widths, these flux
deficits depend only weakly on the detector's resolution, which will
make this easier to detect in practice.  If the required precision
can be attained, we note that a quantitative measure of $j$ will
provide a constraint on the properties of high-density matter that
is independent of, e.g., measurements of $R/M$.

Fig.~5 illustrates the effect of $j$ on the flux deficit ratio,
when the other parameters in the problem (including $\nu_{s}$)
remain fixed. Both profiles are normalized by their maximum depths.
In this example, the flux deficit ratio is $\sim 36$\% greater for
$j=0.188$ ($\nu_{\rm spin} = 400$~Hz) than for $j=0$.

Fig.~6 demonstrates that the double flux minima remain visible
even if other broadening mechanisms, such as Zeeman splitting, make
the intrinsic line width substantially greater than it would
otherwise be. Shown are the predicted profiles of lines with
intrinsic halfwidths of $10^{-3} E_0$ (the same as in the other line
profile calculations reported here) and $5.3 \times 10^{-3} E_0$
(half the halfwidth produced by rotational Doppler broadening). The
latter halfwidth is similar to the magnetic splitting produced by
a $10^9$~G magnetic field, comparable to the field required to
maintain a neutron star with a long-term average accretion rate
$\langle {\dot M} \rangle \sim 0.03 {\dot M}_E$ in spin equilibrium
at the 45~Hz spin rate of \hbox{EXO~0748-676}.

\section{Discussion and Conclusions} \label{sec: 4}

The discovery of line features in the spectrum of
\hbox{EXO~0748-676} by Cottam et al.~(2002) motivated us to explore
the information that can be obtained from analysis of the profiles
of atomic lines formed at the surfaces of neutron stars. We confirm
the finding of Villarreal \& Strohmayer (2004) that the widths of
these features are consistent with formation at the surface of a
star with a spin rate of 45~Hz. We have also shown that stars with
spin rates up to $\sim 600$~Hz can produce lines this narrow if
emission from close to the rotational axis dominates. Confirmation
of these features by future observations, or detection of surface
lines from other LMXBs, will help strengthen and clarify this
picture.

We have demonstrated that the radius-to-mass ratio $R/M$ can be
inferred from spectral line profiles with very little (in general
less than 2\%) systematic error, if the geometric mean of the
energies of the two flux minima (instead of the energy of the
deepest flux minimum), or two line edges is used. Observed line profiles can therefore
be used to constrain EOS models to the 5\% accuracy required to
distinguish between different equations of state (Lattimer \&
Prakash 2001). Finally, we show that neutron star atomic spectral
lines can be used to detect frame-dragging. Such a detection would
require observation of a line profile with two moderately
well-resolved flux minima.

With the discovery of strong evidence for resonance scattering lines
formed at the surface of one neutron star, it can be hoped that
current and future X-ray missions will be able to detect more such
spectral lines. A strong and well-resolved broad line would provide
a wealth of information about the neutron star. One candidate line
is the Fe K$\alpha$ line, given the expected surface temperature and
abundance of iron in accreting sources. Future X-ray spectroscopy
missions will do even better. For example, Constellation-X will have
a resolution and an effective area several times greater than those
of Chandra and XMM. Detection and precise measurement of the
properties of atomic lines formed at the surfaces of neutron stars
would open a new avenue for studying dense matter, neutron stars,
and LMXBs. Simultaneous observations of lines and burst oscillations
by future, even more capable instruments could make possible
phase-resolved line spectroscopy, using the light curve of the burst
oscillation to determine the rotational phase of the star.

\acknowledgments

We thank Arun V. Thampan for providing the code used to compute the
structure of rapidly spinning neutron stars and Geoff Ravenhall for
providing tabulations of the A18$+\delta v+$UIX equation of state.
We also appreciate the particularly useful comments about frame-dragging
provided by the anonymous referee.
This work was supported in part by NSF grant AST~0098436 at
Maryland, by NSF grant AST~0098399 and NASA grants NAG~5-12030 and
NAG~5-8740 at Illinois, and by the funds of the Fortner Endowed
Chair at the University of Illinois.

{}

\clearpage
\begin{figure}
\epsscale{1.0}
\plotone{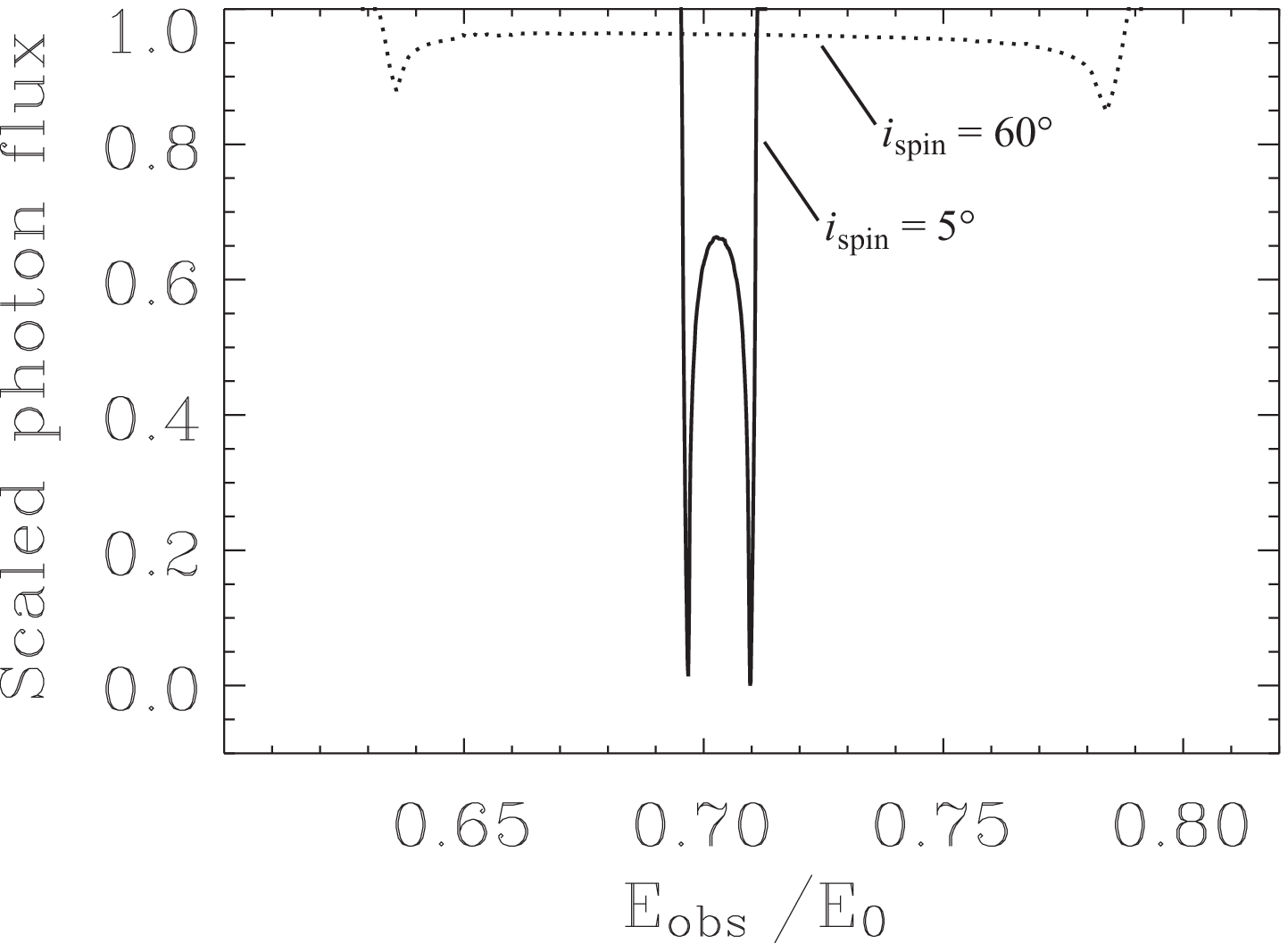}
\caption
{Profiles of a narrow line formed at the stellar surface as seen by
observers at $i_{\rm spin}=5^{\rm o}$ (solid profile) and $i_{\rm spin}=60^{\rm o}$
(dotted profile),
showing that such lines typically have two narrow flux minima.
$E_{\rm obs}/E_{\rm 0}$ is the energy of the observed radiation in
units of the rest energy $E_{\rm 0}$ of the line center. Both
profiles are for EOS model A18$+\delta v+$UIX, $\nu_{s} = 400$~Hz,
$R/M=4.0$, $i_{\rm belt}=90^{\rm o}$, $\Delta i_{\rm belt}=5^{\rm
o}$, and isotropic emission in the frame corotating with the stellar
surface. The $i_{\rm spin}=5^{\rm o}$ profile is scaled in units of
its deepest flux minimum; the $i_{\rm spin}=60^{\rm o}$ profile is
scaled to have the same equivalent width.} \end{figure}

\clearpage
\begin{figure}
\hspace{-4.9 cm}
\epsscale{1.0}
\plotone{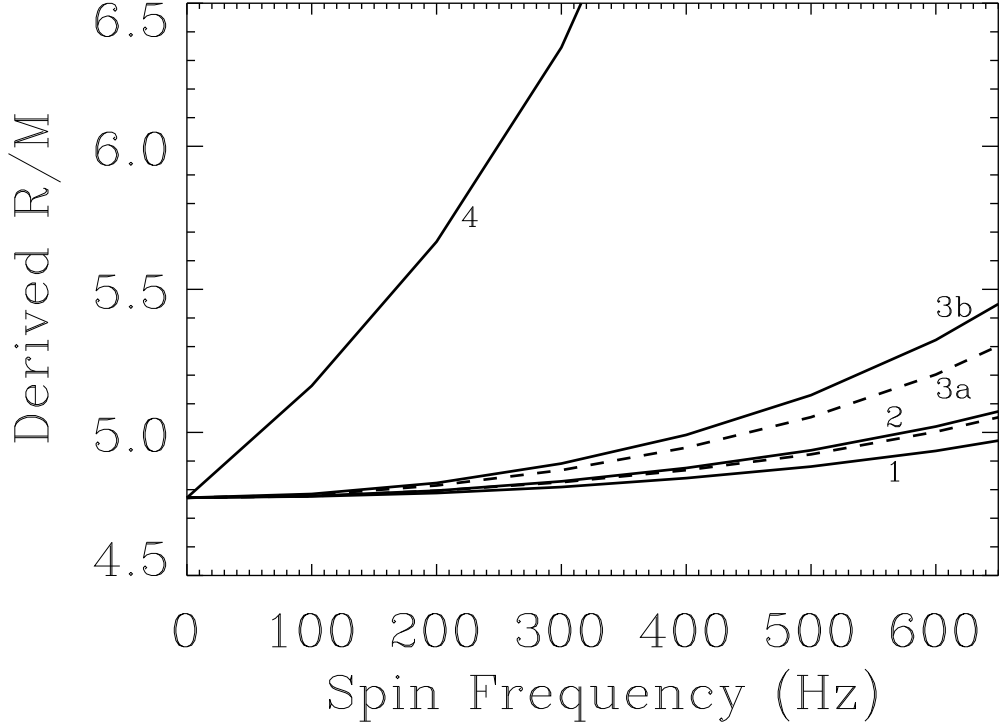}
\vspace{-11.5 cm}
\caption
{Accuracy of $R/M$ estimates using different methods. Curve~1: true
equatorial value of $R/M$ as a function of the stellar spin
frequency. Curve~2: $R/M$ estimates based on the geometric mean of
the energies of the two narrow photon flux minima in the line
profile (dashed curve) and the lower- and higher-energy edges of the
line (solid curve). See text for details. For the parameters assumed
here, these two methods give nearly identical estimates. When these
methods are used, the error in $R/M$ is typically less than 2\%,
even for spin frequencies as high as 600~Hz. Curves~3a and 3b:
estimates based, respectively, on the arithmetic mean of the
energies of the flux minima and the energies of the lower- and
higher-energy edges of the line. Curve~4: estimates based on the
energy of the deepest flux minimum in the line profile. All curves
are for EOS model A18$+\delta v+$UIX, $M=1.6M_{\odot}$, $i_{\rm
spin}= 90^{\rm o}$, $i_{\rm belt}^{\rm p}=60^{\rm o}$, $i_{\rm
belt}^{\rm eq}= 90^{\rm o}$, and isotropic emission in the frame
corotating with the stellar surface.} \end{figure}

\clearpage
\begin{figure}
\hspace{-4.9 cm}
\epsscale{1.0}
\plotone{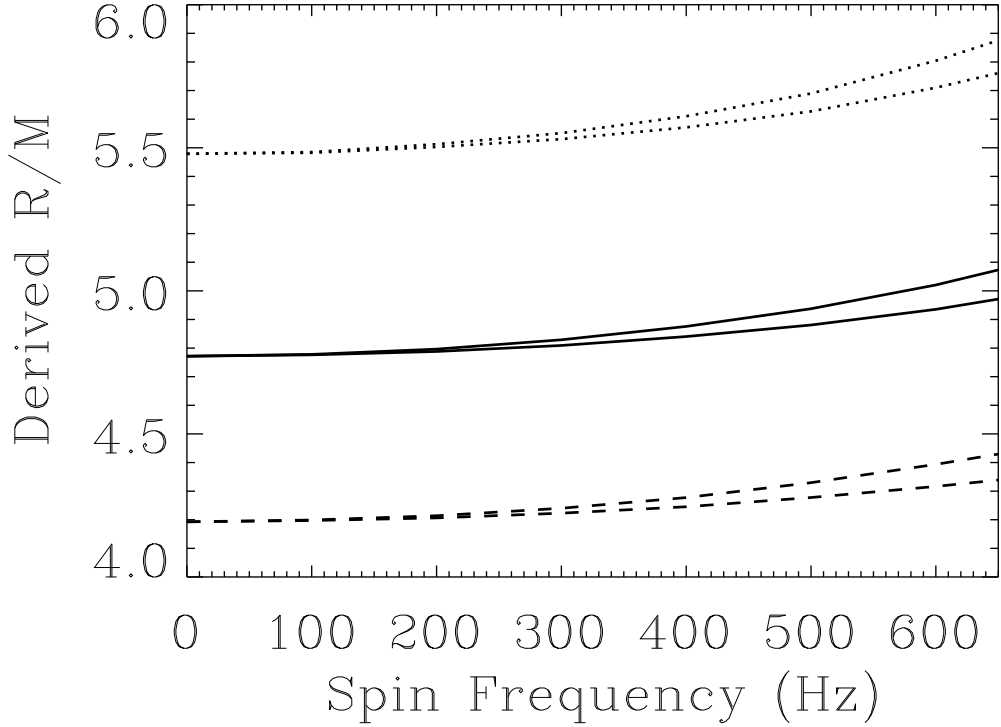}
\vspace{-11.5 cm}
\caption
{Accuracy of $R/M$ estimates for different stellar mass $(M)$ values.
In each of the three sets of curves, the lower curve gives the true equatorial value of $R/M$ as
a function of the stellar spin frequency. The corresponding upper curve gives
the estimate of $R/M$ based on the geometric mean of the energies of the lower- and
higher-energy edges of the line (see text for details). The dotted curves are for
$M = 1.4M_\odot$, the solid curves are for $M = 1.6M_\odot$, and the dashed 
curves are for $M = 1.8M_\odot$. The values of the other parameters are same as
in Fig.~2. The percentage error in $R/M$ estimate very slowly increases with
the increase of $M$ (or, the decrease of $R/M$, when the other independent
parameters remain same).
} \end{figure}

\clearpage
\begin{figure}
\hspace{-4.9 cm}
\epsscale{1.0}
\plotone{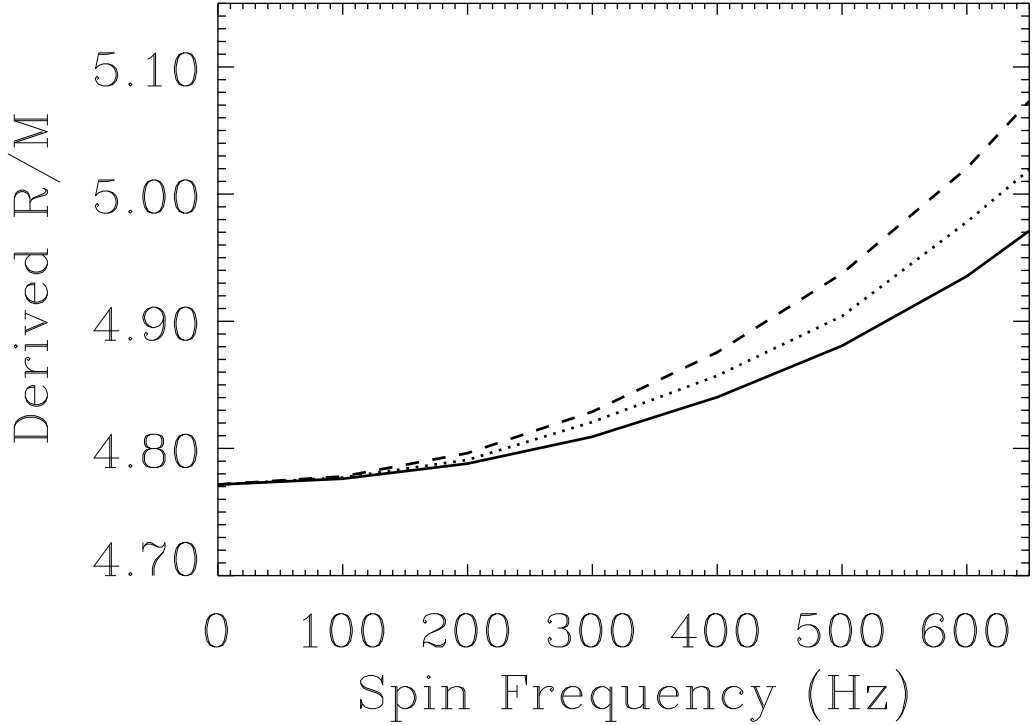}
\vspace{-11.5 cm}
\caption
{Accuracy of $R/M$ estimates for different belt locations and belt widths.
The solid curve gives the true equatorial value of $R/M$ for EOS model
A18$+\delta v+$UIX and $M=1.6M_{\odot}$. 
The geometric means of the energies of the lower- and  higher-energy edges
of the line are the same for the following three belt locations and
widths: (i) belt covering the whole stellar surface, (ii) $i_{\rm belt}^{\rm
p}=60^{\rm o}$ and $i_{\rm belt}^{\rm eq}= 90^{\rm o}$, and (iii) $i_{\rm
belt}=90^{\rm o}$, $\Delta i_{\rm belt}=5^{\rm o}$.
The dashed curve gives the estimate of $R/M$ based on these mean energies.
The dotted curve gives the estimate of $R/M$ for 
$i_{\rm belt}=30^{\rm o}$, $\Delta i_{\rm belt}=5^{\rm o}$.
For all the curves, $i_{\rm spin}= 90^{\rm o}$ and the emission is isotropic
in the frame corotating with the stellar surface.
} \end{figure}

\clearpage
\begin{figure}
\hspace{-4.9 cm}
\epsscale{1.0}
\plotone{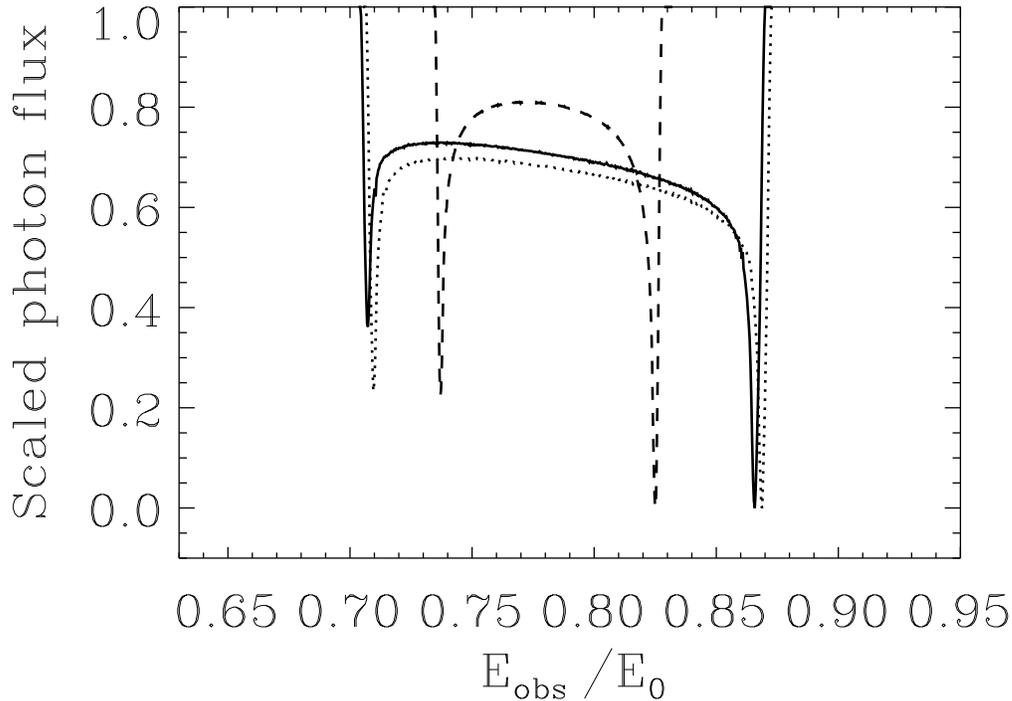}
\vspace{-11.5 cm}
\caption
{Profiles of narrow lines formed at the stellar surface as they
would appear to an observer at $i_{\rm spin}=60^{\rm o}$ if $j=0.0$ (solid profile)
and if $j=0.188$ (dotted profile), showing that measurements of the depths of the
flux minima could be used to detect the general relativistic
dragging of inertial frames. $E_{\rm obs}/E_{0}$ is the energy of
the observed radiation in units of the rest energy of the line. For
each of the profiles the 
flux has been scaled to have a minimum value of 0 and a maximum
value of 1. Both profiles assume EOS model A18$+\delta v+$UIX,
$M=1.5$\msun, $i_{\rm belt}=90^{\rm o}$, $\Delta i_{\rm belt}=5^{\rm
o}$, and isotropic emission in the frame corotating with the stellar
surface. For this EOS and mass, $j=0.188$ for $\nu_{s} = 400$~Hz. A third profile (dashed; 
for $i_{\rm spin}=30^{\rm o}$ and $j=0.0$, all other parameter values are same, 
scaled in the same way as the other two profiles)
is displayed to show that, although 
the frame-dragging and a lower value of $i_{\rm spin}$ have the similar effects on
the depths of the flux minima, these two causes can be distinguished by the
width of the line profile.}
\end{figure}

\clearpage
\begin{figure}
\hspace{-0.4 cm}
\epsscale{1.0}
\plotone{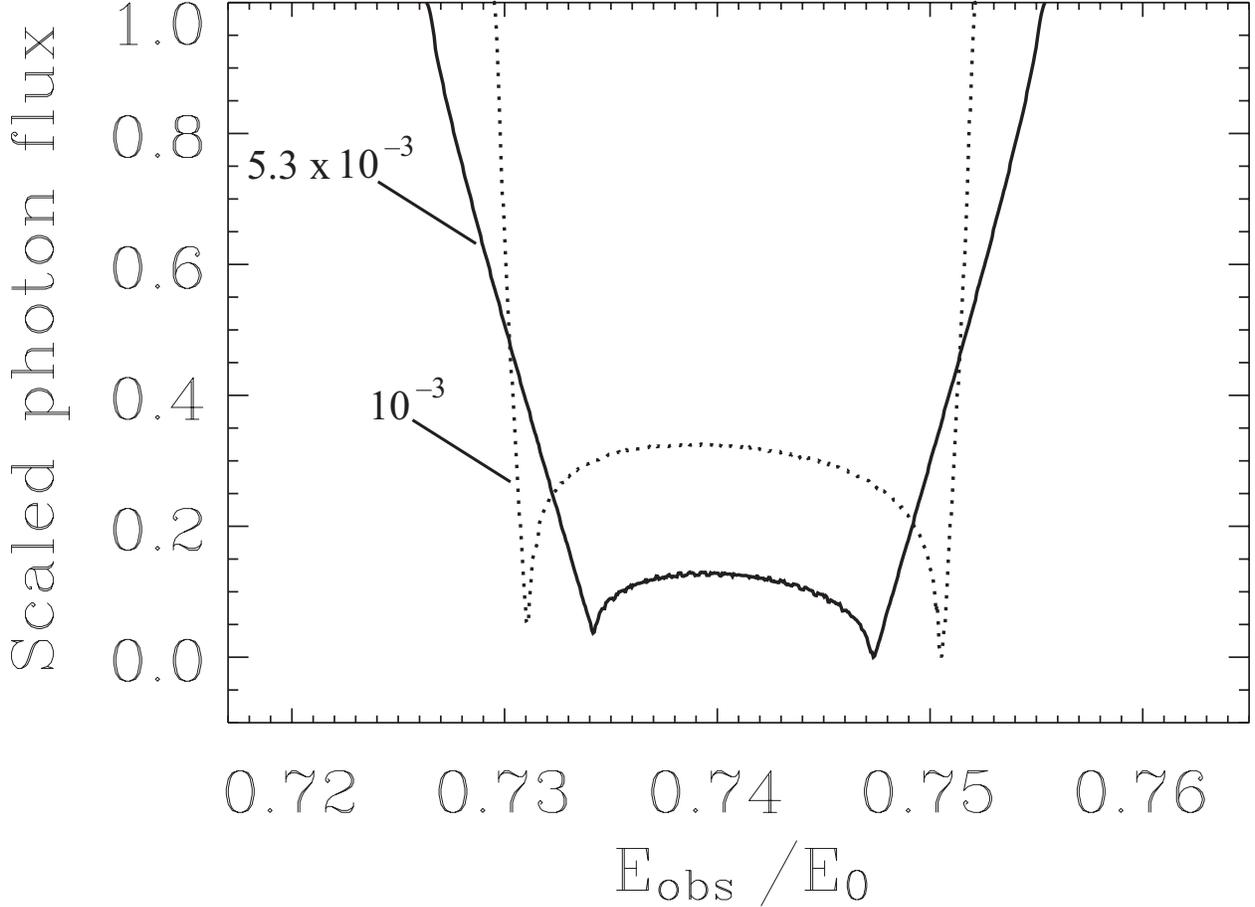}
\vspace{1.0 cm}
\caption
{Effect on line profiles of additional broadening, such as that
produced by unresolved Zeeman splitting. Shown are the observed
profiles of lines with intrinsic halfwidths of $10^{-3} E_0$ (dotted curve; the
same as in the other line profile calculations reported here) and
$5.3 \times 10^{-3} E_0$ (solid curve; half the halfwidth produced by rotational
Doppler broadening, comparable to the splitting produced by a
$10^9$~G magnetic field).  For each of the profiles the flux has
been scaled in units of its deepest minimum.  Both profiles are for
$R/M = 4.43$, $R = 11.5$~km, $\nu_{s} = 45$~Hz, $i_{\rm
belt}=90^{\rm o}$, $\Delta i_{\rm belt}=5^{\rm o}$, isotropic
emission in the frame corotating with the stellar surface, and
$i_{\rm spin}=78^{\rm o}$.}
\end{figure}

\end{document}